\documentclass[preprint,showpacs]{revtex4}

\begin{document}

\title{Topological low-temperature limit of Z(2) spin-gauge theory in three dimensions}
\author{Nelson Yokomizo}
\email{yokomizo@fma.if.usp.br}
\author{Paulo Teotonio-Sobrinho}
\email{teotonio@fma.if.usp.br}
\author{Jo\~ao C. A. Barata}
\email{jbarata@fma.if.usp.br}
\affiliation{Instituto de F\'{i}sica, Universidade de S\~{a}o Paulo, C.P. 66318, 05315-970 S\~{a}o Paulo-SP, Brazil}
\date{\today}

\begin{abstract}

{We study Z(2) lattice gauge theory on triangulations of a compact 3-manifold. We reformulate the theory algebraically, describing it in terms of the structure constants of a bidimensional vector space $\mathcal{H}$ equipped with algebra and coalgebra structures, and prove that in the low-temperature limit $\mathcal{H}$ reduces to a Hopf Algebra, in which case the theory becomes equivalent to a topological field theory. The degeneracy of the ground state is shown to be a topological invariant. This fact is used to compute the zeroth- and first-order terms in the low-temperature expansion of $Z$ for arbitrary triangulations. In finite temperatures, the algebraic reformulation gives rise to new duality relations among classical spin models, related to changes of basis of $\mathcal{H}$.}

\end{abstract}

\maketitle

The Z(2) spin-gauge theory we consider in this work is a lattice gauge theory, locally invariant under the abelian discrete group Z(2), and one of the simplest examples of a gauge theory on the lattice, according to Wilson's ideas \cite{wilson}. There are two main motivations for the study of this model. First, it is hoped that the understanding of this simple example will bring some insight into the physics of more realistic gauge theories. Second, it is known that in three dimensions this theory is dual to the 3D Ising model \cite{wegner}, an outstanding problem in statistical mechanics.

We study Z(2) pure gauge theory on arbitrary finite triangulations of a compact 3-manifold. Usually, one considers cubic lattices \cite{itzykson,kogut}, but we do not follow this procedure here, in order to achieve a more general framework for the study of topological properties. It is always supposed that some kind of continuum limit is to be taken, after all, in which the details of the lattice become unimportant, and so there should be no problem in taking alternative discretisations. 

The theory is defined as follows. Let $\sigma_a=(-1)^a,\; a=0,1$, be gauge variables sitting at the links $a$ of a lattice $L$. Let $\omega(f)$ be the product of all gauge variables at the boundary of a particular face $f$. Then the action is $S = \beta \sum_f \omega(f)$, where $\beta$ is the coupling constant of the theory, and the sum runs over all faces $f$ of the lattice $L$. The partition function is given by $Z=\sum_{\{\sigma_a\}} \exp(S)$, the sum running over all gauge configurations $\{\sigma_a\}$. We will find it convenient to rewrite $Z$ as a product of local Boltzmann weights, in the form
\begin{equation}
Z=\sum_{\{\sigma_a\}} \prod_f M(f), \label{eq:Z2}
\end{equation}
where $M(f)=\exp[\beta \omega(f)]$. The low-temperature limit is obtained by letting $\beta$ go to infinity.

As the lattice $L$ is a triangulation, all its faces are triangular. Thus the weights $M(f)$ describe a three-spin interaction. We write them as $M_{abc}$, where $a,b,c$, are the indices of the gauge variables $\sigma_a$ at the boundary of $f$. It follows that $M_{abc}=\exp[(-1)^{a+b+c}\, \beta]$. 

We give the theory an algebraic interpretation, in terms of a bidimensional algebra $\mathcal{H}$. For that, we follow prescriptions introduced by Chung, Fukuma and Shapere (CFS) in \cite{cfs}, in the context of topological quantum field theory \cite{tqft}. There, it is shown how to build up a field theory on a three-dimensional lattice from the knowledge of the structure constants of an algebra of interest. Such procedure allows one to encode symmetries of the theory in symmetries of the algebra, and was first applied in the investigation of topological invariance. The main result found was that the lattice theory is topological whenever $\mathcal{H}$ is a Hopf algebra \cite{cfs,kuperberg}.

Here we show that Z(2) pure gauge theory can be reformulated as a CFS theory. We display explicitly an algebra $\mathcal{H}$ which leads to Z(2) gauge theory through the use of the prescriptions of \cite{cfs}. The algebra $\mathcal{H}$ depends on the value of $\beta$. In the low-temperature limit, the Hopf algebra axioms hold, up to the appearance of some extra factors in its defining relations. It follows that the theory is almost topological in this limit. In fact, we prove that $Z|_{\beta \rightarrow \infty}$ is essentially the product of a volume-dependent factor and a topological invariant, giving an example of a 3D quasitopological field theory \cite{teot}. This decomposition leads to a solution of the leading terms in the low-temperature expansion of $Z$ described in \cite{expansion}. The zeroth-order term depends only on the degeneracy $\Xi(L)$ of the ground state. We prove that it is a topological invariant, and show how to compute it. The first-order correction depends on simple combinatorial factors. It is hoped that higher order terms can be dealt with in a similar fashion.

Let us describe the algebra $\mathcal{H}$. Let $B=\{\phi_0,\phi_1\}$ be a basis of $\mathcal{H}$, and $B^\star = \{\psi^0,\psi^1\}$ its dual basis, for which $\psi^a(\phi_b)=\delta^a_b$. We define a linear product $M:\mathcal{H} \otimes \mathcal{H} \mapsto \mathcal{H}$, which sends the the pair of vectors $u,v \in \mathcal{H}$ into $u \cdot v \in \mathcal{H}$, as usual. The products of the basis vectors read
\begin{equation}
\begin{array}{ccc}
\phi_0 \cdot \phi_0 = &\phi_1 \cdot \phi_1 = &f \; \phi_0 + e \; \phi_1, \\
\phi_0 \cdot \phi_1 = &\phi_1 \cdot \phi_0 = &e \; \phi_0 + f \; \phi_1, \label{eq:prod}
\end{array}
\end{equation} 
where $ e = \rho^{-1} \sinh(x)$, $f= \rho^{-1} \cosh(x)$, and $\rho$ and $x$ are fixed by the coupling constant $\beta$ through
\begin{equation}
\begin{array}{cc}
e^{-2 \beta}  = &\tanh(3x), \\
\rho^{-6}  = &2 \sinh(6x). \label{eq:beta-x}
\end{array}
\end{equation}
The algebra has an unity $\iota = \rho \, \cosh(x) \, \phi_0 - \rho \, \sinh(x) \, \phi_1$, and is associative. Its structure constants are the coefficients $M_{ab}^c=\psi^c(\phi_a \cdot \phi_b)$ of the product tensor $M$.

Definining the trace $T \in \mathcal{H}^\star$ as the dual vector with coefficients $T_a=M_{ab}^b$, it follows that $T=2f\psi^0 + 2e \psi^1$, and that
\begin{equation}
\begin{array}{ccc}
M_{abc} &=& T(\phi_a \cdot \phi_b \cdot \phi_c) \\
&=& M_{ab}^x M_{xc}^y M_{yz}^z. \label{eq:weights}
\end{array}
\end{equation}
This formula shows how the weights of Z(2) gauge theory are related to the structure constants of the algebra $\mathcal{H}$: they are coefficients of a three-indexed covariant tensor  built from the product tensor $M$ alone.

The expression in Eq. \ref{eq:weights} is an example of the CFS formalism, and the first step in our reformulation of the Z(2) theory. In this formalism, the weights $M_{abc}$ are thought as determined by the algebra $\mathcal{H}$. If a different algebra was taken, there would be different local weights $M_{abc}$ at the lattice faces, and another set of configurations $a$ at the links. Thus, a different lattice field theory, determined by the structure constants of the algebra. For the particular algebra $\mathcal{H}$ and basis $\mathcal{B}$ we defined, the formula happens to give the weights $M_{abc}=\exp[(-1)^{a+b+c}\, \beta]$ of Z(2) pure gauge theory, as can be checked. The problem of finding an algebra $\mathcal{H}$ which gives a specified set of weights $M_{abc}$ may not be a trivial task. In our case it involved the solution of a system with eight equations, reduced to a single ninth degree equation in two variables.

We carry the algebraic reformulation further. We want to write the partition function $Z$ in tensorial form. As it is a number, it shall be a scalar of $\mathcal{H}$. Therefore, some kind of contravariant tensor is needed, so that we can build contractions with $M_{abc}$ in order to define scalars. For that purpose, we give $\mathcal{H}$ a coalgebra structure, to whose definition we now turn.

Let $\Delta: \mathcal{H} \mapsto \mathcal{H} \otimes \mathcal{H}$ be a coproduct on $\mathcal{H}$, whose action on basis vectors is given by
\begin{equation}
\Delta(\phi_a)=\phi_a \otimes \phi_a. \label{eq:coprod}
\end{equation}
The coalgebra so defined has a counity $\varepsilon=\psi^0 + \psi^1$, and is coassociative. Its structure constants are the coefficients $\Delta_a^{bc}=(\psi^b \otimes \psi^c)(\Delta(\phi_a))=\delta_a^b \, \delta_a^c$ of the tensor $\Delta$.

In analogy with Eq. \ref{eq:weights}, we define the $n$-indexed contravariant tensors $\Delta^{a_1 a_2 \dots a_{n-1} a_n}$ by
\begin{equation}
\Delta^{a_1 a_2 \dots a_{n-1} a_n} = \Delta_{x_1}^{x_1 x_2} \, \Delta_{x_2}^{a_1 x_3} \, \Delta_{x_3}^{a_2 x_4} \dots \, \Delta_{x_n}^{a_{n-1} a_n}. \label{eq:hinges}
\end{equation}
For the coalgebra defined by Eq. \ref{eq:coprod}, they read
\begin{equation}
\Delta^{a_1 a_2 \dots a_{n-1} a_n} = \delta_{a_1,a_n} \delta_{a_2,a_n} \dots \delta_{a_{n-1},a_n}, \label{eq:links}
\end{equation}
where the $\delta's$ are Kronecker deltas.

The expression in Eq. \ref{eq:hinges} is another formula of the CFS formalism. It defines local weights which are assigned to links. In their formalism, the local configurations are not assigned to links or vertices, as usual. Instead, if a link $a$ is shared by $n$ faces, then there are $n$ configurations $a_i,i=1,2,\dots,n$, sitting at the intersections of the link and the faces. The link is thought of not as a thin line, but as a sort of 'hinge': an object with $n$ strips emerging from a central line, where faces are glued (see Ref. \cite{cfs}). There is a configuration $a_i$ at each strip. The coefficients $\Delta^{a_1 a_2 \dots a_{n-1} a_n}$ are local weights assigned to hinges, determined by the configurations at its strips. Different coalgebras lead to different hinge weights and, therefore, to distinct lattice field theories.

The partition function is defined as a scalar written with contractions of indices of hinge and face weights. Consider the outer product of all hinge and face weights in the lattice $L$. For each hinge strip, there is a configuration $a$, which appears as a lower index in a face weight, and as an upper index in some hinge weight. We contract this pair of indices. Then we do the same for all hinge strips. A scalar is obtained, which is defined as the partition function $Z$ of the theory. In the original work of CFS, there is also an antipode operator $S:\mathcal{H} \mapsto \mathcal{H}$, which interfere in the contractions, and is used to define orientations. The rule we gave here is equivalent to setting $S$ equal to the identity, $S_a^b = \delta_a^b$. This special case is enough for our purposes.

Now it just remains to note that the special form of the coproduct given in Eq. \ref{eq:coprod} is just the one needed in order to recover the interpretation of configurations at links. For, according to Eq. \ref{eq:links}, in this case the hinge weights are zero, unless all its indices are equal. But then this index can be thought of as assigned to the link itself. Therefore, under such condition, a global configuration is an assignment of an index $a=0,1$ to each link of $L$. There is an weight $\prod_f M_{abc}(f)$ for each configuration. The contractions of indices which define $Z$ automatically implements the sum over configurations. For $\mathcal{H}$, the weights are simply $M_{abc}=\exp[(-1)^{a+b+c}\, \beta]$, and the result is Z(2) pure gauge theory's partition function, as written in Eq. \ref{eq:Z2}. That completes our reformulation of the theory.

Let us turn to the low-temperature limit of the theory now. It is obtained by letting $\beta\rightarrow \infty$, or, equivalently, $x \rightarrow 0$, as shown by Eq. \ref{eq:beta-x}. The partition function diverges in this limit, $\lim_{x\rightarrow 0}Z=\infty$, and we want to understand how that happens, looking at the algebraic reformulation of the theory. The structure constants of $\mathcal{H}$ are not both well-defined when $x$ goes to zero. The coefficient $e$ presents no problem, as $\lim_{x\rightarrow0} e(x) = 0$, but for $f$ we have $\lim_{x\rightarrow 0} \, f(x) = \infty$. We will show that the divergences of the algebra coefficients are related to the divergence of the partition function itself in this limit. Such interpretation comes from a simple algebraic manipulation. A continuous change of basis of $\mathcal{H}$, parametrised by $x$, is used to isolate the singular part of $Z$, which is then found to be a function of $f$.

Note that the partition function $Z$, being a scalar of $\mathcal{H}$, does not change its value if a different basis $B^\prime$ is chosen to write the link and face weights of the theory. But the weights itself do change, and thus we have a different lattice field theory with the same partition function. For instance, consider the change of basis, which we will use in the low-temperature limit, given by the scaling transformation
\begin{equation}
\begin{array}{c}
\xi_0 = f^{-1} \phi_0, \\
\xi_1 = f^{-1} \phi_1. \label{eq:scaling}
\end{array}
\end{equation}
In the new basis $B^\prime = \{\xi_0,\xi_1\}$, the structure constants of $\mathcal{H}$ are given by
\begin{equation}
\begin{array}{ccc}
\xi_0 \cdot \xi_0 = &\xi_1 \cdot \xi_1 = & \xi_0 + \frac{e}{f} \; \phi_1, \\
\xi_0 \cdot \xi_1 = &\xi_1 \cdot \xi_0 = &\frac{e}{f} \; \phi_0 + \xi_1, \label{eq:newprod}
\end{array}
\end{equation} 
and the coproduct is
\begin{equation}
\Delta(\xi_a)=f \, \xi_a \otimes \xi_a.
\end{equation}
The hinge weights, determined by Eq. \ref{eq:hinges}, are now given by
\begin{equation}
\Delta^{\prime a_1 a_2 \dots a_{n-1} a_n} = f^n \, \delta_{a_1,a_n} \delta_{a_2,a_n} \dots \delta_{a_{n-1},a_n}.
\end{equation}
They still allow a link configuration interpretation of the theory. But now there is a local weight $f^n$ at each link, which does not depend on the configuration at the link. The face weights given by Eq. \ref{eq:weights} read
\begin{equation}
\begin{array}{cc}
M^\prime_{000} = M^\prime_{011} =& 2 + 3(\frac{e}{f})^2 , \\
M^\prime_{001} = M^\prime_{111} =& 6 \frac{e}{f} + 2 (\frac{e}{f})^3, \label{eq:newbasis}
\end{array}
\end{equation}
and are cyclically symmetric. 

The theory obtained with the basis $B^\prime$ can thus be summarised as follows. There are configurations $a=0,1$ at all lattice links. At each link, there is a local weight $f^n$, where $n$ is the number of faces meeting at the link. On each face, there is a cyclically symmetric weight whose value is given by Eq. \ref{eq:newbasis}. The partition function $Z$ is the sum over all link configurations of the product of all face weights in the lattice. This partition function has the same value as that of Z(2) pure gauge theory. This is an example of a new duality relation between lattice spin models, and we will discuss it later in this paper.

A simplification can be achieved if the contribution of the weights $f^n$ is removed from $Z$. For any configuration, these weights taken together lead to a factor $f^{3N_f}$, where $N_f$ is the total number of faces in the lattice. Thus if we define a new theory with face weights given by Eq. \ref{eq:newbasis}, but with no link weights, it will have a partition function $X=Z f^{-3N_f}$. Its CFS description is given by the algebra displayed in Eq. \ref{eq:newprod}, with the coproduct of Eq. \ref{eq:coprod}. 

We consider the low-temperature limit of such theory. This is enough for us, since a solution for $X$ gives a solution for $Z$. But when $x\rightarrow 0 \; (\beta\rightarrow \infty)$, the structure constants are:
\begin{equation}
\begin{array}{c}
\xi_0 \cdot \xi_0 = \xi_1 \cdot \xi_1 =  \xi_0,  \\
\xi_0 \cdot \xi_1 = \xi_1 \cdot \xi_0 =  \xi_1, \\
\Delta(\xi_a) = \xi_a \otimes \xi_a. \label{eq:lowalg}
\end{array}
\end{equation}
These product and coproduct, taken together with an antipode $S^a_b=\delta^a_b$, constitute a Hopf algebra. Thus the partition function $X|_{\beta \rightarrow \infty}$ is a topological invariant: its value does not depend on the details of the lattice, being determined by the topology of $L$ alone. 

Actually, as we defined the theory, there is an extra factor yet. From Kuperberg's work \cite{kuperberg}, we have that $X|_{\beta \rightarrow \infty} = \Xi(L) 2^{2 N_t} 2^{N_v-1}$, where $\Xi(L)$ is Kuperberg's topological invariant, and $N_t$ and $N_v$ are the total number of tetrahedra and vertices in $L$, respectively \footnote{Kuperberg's theory is equivalent to CFS theory, but is defined on Heegard diagrams $D$, instead of on lattices. For any Hopf algebra $\mathcal{H}$, the partition function is $Z(\mathcal{H})=\#(\mathcal{D},\mathcal{H})(\dim \mathcal{H})^{n_u + n_v - g(S)}$, where $\#(\mathcal{D},\mathcal{H})$ is a topological invariant, $g$ is the genus of $\mathcal{D}$, and $n_u,n_v$ are the number of upper and lower curves of $\mathcal{D}$, respectively. In the language of triangulations, this formula reduces to $Z(\mathcal{H}) = \Xi(L,\mathcal{H}) 2^{2 N_t} 2^{N_v-1}$, where $\Xi(L,\mathcal{H})=\#(\mathcal{D},\mathcal{H})$.}. Noting that $N_f=2N_t$ for any triangulation, we are finally able to write the partition function of Z(2) lattice gauge theory in the low-temperature limit as
\begin{equation}
Z|_{\beta \rightarrow \infty} =\Xi(L) (2f^3)^{2 N_t} 2^{N_v-1}, \label{eq:lowsol}
\end{equation}
i.e., as the product of the topological invariant $\Xi(L)$, a volume-dependent factor $(2f^3)^{2 N_t}$, and a factor $2^{N_v-1}$ which counts gauge equivalent configurations. $\Xi(L)$ depends on the topology of the lattice only. It is the partition function of a Kuperberg model \cite{kuperberg} defined with the low-temperature algebra of $X$ as given in Eq. \ref{eq:lowalg}. The factor $(2f^3)^{2 N_t}$ gives the temperature dependence of $Z$ as $\beta$ approaches $\infty$. This factor diverges at zero temperature, and is the singular part of $Z|_{\beta \rightarrow \infty}$. For a given $\beta$, its value depends only on $N_t$, the discrete version of the volume. The remaining factor $2^{N_v-1}$ is quite irrelevant, and could well be eliminated by gauge fixing.

The formula in Eq. \ref{eq:lowsol}, in addition to the algebraic reformulation of the theory, is the main result of this paper. Next we will discuss it further, and show how to use the equation in explicit calculations. In particular, we give solutions for the zeroth and first-order terms in the low-temperature expansion of $Z$.

To begin, we give an interpretation for the topological invariant $\Xi(L)$. For that, we evaluate the low-temperature limit of the weights $M_{abc}^\prime$ which define $X$. As $\lim_{x \rightarrow 0} (e/f)=0$, then we can write
\begin{equation}
\begin{array}{cc}
M^\prime_{000}(=M^\prime_{011})|_{{\beta \rightarrow \infty}} =2, \\
M^\prime_{001}(=M^\prime_{111})|_{{\beta \rightarrow \infty}} =0. \label{lowlimit}
\end{array} \;
\end{equation}
These formulae have a very simple meaning. The weight of a gauge configuration is the product of all local weights $M_{abc}^\prime$. According to Eq. \ref{lowlimit}, it is non-zero, in the low-temperature limit, only if $w(f)=1$ at all faces of $L$. In this case, it is equal to $2^{N_f}$. Suppose there are $A_0$ such gauge configurations. Then $X|_{\beta \rightarrow \infty} =  2^{N_f} A_0$, and so
\begin{equation}
\Xi(L) = \frac{A_0}{2^{N_v -1}}, \label{eq:xi}
\end{equation}
i.e., $\Xi(L)$ is the number of equivalence classes of gauge configurations for which $w(f)=1,\forall f \in L$. This condition selects gauge configurations with minimal energy, that is, ground state configurations. Thus $\Xi(L)$ is the degeneracy of the ground state on the lattice $L$, and we have just proved that it is a topological invariant.  The determination of this number is the first step in any low-temperature expansion of the theory, and is severely simplified due to topological symmetry. The evaluation of $A_0$ is naturally a hard combinatorial problem. One must enumerate all gauge configurations on $L$, check one by one if the ground state condition is satisfied, and count the number of occurrences. We give an alternative procedure. There is a topological field theory which performs this counting. Its partition function is $X|_{\beta \rightarrow \infty} = 2^{N_f} A_0$, and one can take the simplest lattice $L^\prime$ homeomorphic to $L$ in order to evaluate it, as topological symmetry implies that $X(L)=X(L^\prime)$. For the usual topologies, $L^\prime$ will be a lattice with a small number of tetrahedra, and the calculation of $X|_{\beta \rightarrow \infty}$ is feasible (a small finite sum). Then we get a solution for $A_0$, namely, $A_0 = 2^{-N_f} X|_{\beta \rightarrow \infty}$.

Now consider the low-temperature expansion of $Z$ given in \cite{expansion}. At zero temperature, only ground state configurations are accessible. These configurations give the zeroth-order approximation of $Z$, which reads
\begin{equation}
\begin{array}{ccc}
Z^{(0)} &\equiv& A_0 e^{\beta N_f} \\
& =& 2^{-N_f} e^{2 \beta N_t} X|_{\beta \rightarrow \infty}.
\end{array}
\end{equation}
The first-order correction can also be solved. It consists of contributions coming from elementary excitations of the lattice. Pick a ground state configuration. There is an weight $e^\beta$ at each face. Now let $a$ be any link, and $\sigma_a$ the local gauge variable at it. Invert its sign, $\sigma_a \mapsto -\sigma_a$. A new gauge configuration is obtained, for which $w(f)=1$ at all faces, except for those meeting $a$. The contributions for the first-order correction of $Z$ come from these configurations. We write it as
\begin{equation}
Z^{(1)} = A_0 \sum_a e^{\beta (N_f - 2 n_a)},
\end{equation}
where $n_a$ is the number of faces meeting the link $a$. The determination of the combinatorial factors $n_a$ should not bring problems. For instance, for a cubic lattice, $n_a=4$ is constant, and we have
\begin{equation}
Z^{(1)} = 2^{-N_f} N_l \, e^{\beta (N_f - 8)} \, X|_{\beta \rightarrow \infty}.
\end{equation}
This is our solution for $Z^{(1)}$, and another example of a calculation based on the formalism we presented here.

Before concluding the work, we would like to get back to Eqs. \ref{eq:scaling}-\ref{eq:newbasis}, and discuss the meaning of the change of basis presented there. The theory given by the original basis $\mathcal{B}$ is just Z(2) gauge theory, which can be thought of as a spin model. The spins $\sigma_a$ are situated at the links of $L$, and there are Boltzmann local weights $M_{abc}$ at faces. After the change of basis, a new spin model is defined. The spins are still at the lattice links, but the face weights are $M^\prime_{abc}$, and there are also link weights. Yet, the partition function is the same. Thus we have an example of a new duality between spin models, similar to those depicted in \cite{wegner}. In the CFS formalism, there is one such duality for each change of basis of $\mathcal{H}$. If, beyond the scaling transformations studied in this work, more general changes of basis are considered, one may hope that a larger class of new dualities will arise. This method for studying dualities can be applied to any lattice theory which can be reformulated as a CFS theory.

To sum up, we have shown in this paper that Z(2) lattice gauge theory can be reformulated as a CFS theory, giving it an algebraic interpretation in terms of a vector space $\mathcal{H}$ with algebra and coalgebra structure, and proved that the low-temperature limit of the theory is equivalent to a topological field theory. The reformulation is based on the application of mathematical prescriptions coming from topological field theories to the realm of lattice gauge theory. Calculations made in the low-temperature limit are considerably simplified when topological symmetry is taken into account, and can be evaluated for arbitrary triangulations. Moreover, we have found that any change of basis of $\mathcal{H}$ is related to some duality relation among spin models. This observation leads to an algebraic method for the investigation of dualities, which we shall develop elsewhere.

We would like to thank A. P. Balachandran for his helpful comments and interest. N. Y. acknowledges the Physics Institute at USP for the kind hospitality.


\begin{thebibliography}{10}
\bibitem{wilson} K. Wilson, Phys. Rev. D \textbf{10}, 2445 (1974)
\bibitem{wegner} F. J. Wegner, J. Math. Phys. \textbf{12}, 2259 (1971)
\bibitem{itzykson} R. Balian, J. M. Drouffe, and C. Itzykson, Phys. Rev. D \textbf{11}, 2098 (1975)
\bibitem{kogut} J. B. Kogut, Rev. Mod. Phys. \textbf{51}, 659 (1979)
\bibitem{cfs} S. Chung, M. Fukuma, and A. Shapere, Int. J. Mod. Phys. A \textbf{9}, 1305 (1994)
\bibitem{tqft} L. Crane, Comm. Math. Phys. \textbf{135}, 615 (1991)
\bibitem{kuperberg} G. Kuperberg, Int. J. Math. \textbf{2}, 41 (1991)
\bibitem{teot} P. Teotonio-Sobrinho and B. G. C. Cunha, Int. J. Mod. Phys. A \textbf{13}, 3667 (1998)
\bibitem{expansion} R. Marra, and S. Miracle-Sol\'e, Commun. Math. Phys. \textbf{67}, 233 (1979) 
\end{thebibliography}
\end{document}